# Big Crunch-based omnidirectional light concentrators

Igor I. Smolyaninov, Yu-Ju Hung

*Department of Electrical and Computer Engineering, University of Maryland, College Park, MD 20742, USA*

**Omnidirectional light concentration remains an unsolved problem despite such important practical applications as design of efficient mobile photovoltaic cells. Optical black hole designs developed recently offer partial solution to this problem. However, even these solutions are not truly omnidirectional since they do not exhibit a horizon, and at large enough incidence angles light may be trapped into quasi-stationary orbits around such imperfect optical black holes. Here we propose and realize experimentally another gravity-inspired design of a broadband omnidirectional light concentrator based on the cosmological Big Crunch solutions. By mimicking the Big Crunch spacetime via corresponding effective optical metric we make sure that every photon world line terminates in a single point.**

Concentrated solar power systems use various optical elements to collect solar energy over large area in various portions of electromagnetic spectrum and concentrate it into a much smaller area, where it is converted into other forms of energy, such as electric power. Despite large practical importance of this task, omnidirectional light concentration remains an unsolved problem. Various conventional solutions become more difficult to implement in case of small mobile photovoltaic cells, which led to recent development of the optical "black hole" concept [1-4]. The goal of optical black hole designs is to make light collection as efficient as possible by making light collectors compact and omni-directional. However, even these solutions are not truly omnidirectional since they do not exhibit a true horizon, and at large enough incidence angles light may be trapped into quasi-stationary orbits around such imperfect optical black holes. Nevertheless, this area of research appears to be quite promising and exciting. It is based on the recently developed field of transformation optics [5,6] and understanding of its close ties with general relativity [7]. Based on this understanding, here we propose another gravity-inspired design of a broadband omnidirectional light concentrator, which emulates the cosmological Big Crunch solutions. By mimicking the Big Crunch spacetime (Fig.1) via corresponding effective optical metric we make sure that every photon world line terminates in a single point. Obviously, such a design (if made sufficiently broadband) leads to the most efficient light concentrator possible.

Our design uses unusual electromagnetic properties of hyperbolic metamaterials. These materials are able to guide and manipulate electromagnetic fields on a spatial scale much smaller than the free space wavelength [8-13]. Recently it was demonstrated that wave equation describing propagation of extraordinary light inside hyperbolic metamaterials exhibits 2+1 dimensional Lorentz symmetry. The role of time in the corresponding effective 3D Minkowski spacetime is played by the spatial coordinate aligned with the local direction of optical axis of the metamaterial [14,15]. As a result, light rays inside a hyperbolic metamaterial behave as world lines of free particles propagating through a corresponding 2+1 dimensional spacetime. If the Big Crunch metric is emulated, all the possible light rays concentrate in one point.

The spacetime metric of the cosmological Big Crunch solution may be obtained as a particular case of Friedmann–Lemaître–Robertson–Walker (FLRW) metric

$$ds^2 = -c^2 dt^2 + a(t) d\Sigma^2 \qquad (1)$$

where $d\Sigma^2$ represents spatial element of a 3-dimensional space of uniform curvature. $d\Sigma^2$ does not depend on time, so that FLRW metric depends on $t$ only via the "scale factor" $a(t)$. At the Big Crunch point $a(t) \to 0$. The Klein-Gordon equation describing motion of a massive particle in a gravitational field [16]

$$\frac{1}{\sqrt{-g}} \frac{\partial}{\partial x^i}\left(g^{ik}\sqrt{-g}\frac{\partial \varphi}{\partial x^k}\right) = \frac{m^2 c^2}{\hbar^2}\varphi \qquad (2)$$

may be emulated for the FLRW metric using hyperbolic metamaterials as follows. Let us consider the case of a flat two-dimensional $d\Sigma^2$ space, so that

$$ds^2 = -c^2 dt^2 + a(t)\left(dx^2 + dy^2\right), \qquad (3)$$

and the resulting Klein-Gordon equation becomes

$$-\frac{\partial^2 \varphi}{\partial t^2} + \frac{1}{a(t)^2}\left(\frac{\partial^2 \varphi}{\partial x^2} + \frac{\partial^2 \varphi}{\partial y^2}\right) - \frac{2}{a(t)}\frac{\partial a}{\partial t}\frac{\partial \varphi}{\partial t} = \frac{m^2 c^2}{\hbar^2}\varphi \qquad (4)$$

Now let us consider extraordinary photons propagating in a hyperbolic metamaterial having $z$-dependent $\varepsilon_1=\varepsilon_x=\varepsilon_y >0$ and $\varepsilon_2=\varepsilon_z <0$, and assume that this behavior holds in some frequency range around $\omega=\omega_0$. For such photons vector $\vec{E}$ is parallel to the plane defined by the k–vector of the wave and the optical axis. Since hyperbolic metamaterials exhibit large dispersion, we will work in the frequency domain and write the macroscopic Maxwell equations as

$$\frac{\omega^2}{c^2}\vec{D}_\omega = \vec{\nabla}\times\vec{\nabla}\times\vec{E}_\omega \quad \text{and} \quad \vec{D}_\omega = \vec{\vec{\varepsilon}}_\omega \vec{E}_\omega \qquad (5)$$

Let us introduce a photon wave function as $\varphi_\omega = E_{z\omega}$, and assume that the metamaterial is illuminated by coherent CW laser field at frequency $\omega_0$. Let us study spatial distribution of the extraordinary field $\varphi_\omega$ at this frequency. Taking into account $z$ derivatives of $\varepsilon_1$ and $\varepsilon_2$, eq.(5) results in the following equation for $\varphi_\omega$:

$$-\frac{\partial^2 \varphi_\omega}{\varepsilon_1 \partial z^2} + \frac{1}{(-\varepsilon_2)}\left(\frac{\partial^2 \varphi_\omega}{\partial x^2} + \frac{\partial^2 \varphi_\omega}{\partial y^2}\right) + \left(\frac{1}{\varepsilon_1^2}\left(\frac{\partial \varepsilon_1}{\partial z}\right) - \frac{2}{\varepsilon_1 \varepsilon_2}\left(\frac{\partial \varepsilon_2}{\partial z}\right)\right)\left(\frac{\partial \varphi_\omega}{\partial z}\right) +$$

$$+ \frac{\varphi_\omega}{\varepsilon_1 \varepsilon_2}\left(\frac{1}{\varepsilon_1}\left(\frac{\partial \varepsilon_1}{\partial z}\right)\left(\frac{\partial \varepsilon_2}{\partial z}\right) - \left(\frac{\partial^2 \varepsilon_2}{\partial z^2}\right)\right) = \frac{\omega_0^2}{c^2}\varphi_\omega = \frac{m^* c^2}{\hbar^2}\varphi_\omega \quad (6)$$

where $m^*$ is the effective mass. Similar to other hyperbolic metamaterial geometries considered in [14,15], the Maxwell equation (6) resembles the Klein-Gordon equation (4) if we identify $z=\tau$ as a timelike coordinate of the effective 2+1 dimensional "optical spacetime", and introduce an effective scale factor

$$a(z)^2 = a(\tau)^2 = -\frac{\varepsilon_2}{\varepsilon_1} \quad (7)$$

Indeed, if we introduce a new wave function in eq.(4) as $\psi = a\phi$, it becomes

$$-\frac{\partial^2 \psi}{\partial t^2} + \frac{1}{a(t)^2}\left(\frac{\partial^2 \psi}{\partial x^2} + \frac{\partial^2 \psi}{\partial y^2}\right) = \left(\frac{m^2 c^2}{\hbar^2} - \frac{1}{a}\frac{\partial^2 a}{\partial t^2}\right)\psi \quad (8)$$

So it is clear that no particle world line can escape the Big Crunch point at $a(t) \to 0$, since the space itself shrinks to a point. In a similar fashion, if we introduce $\psi = \varepsilon_2 \phi_\omega / \varepsilon_1^{1/2}$, eq.(6) becomes

$$-\frac{\partial^2 \psi}{\partial z^2} + \frac{\varepsilon_1}{(-\varepsilon_2)}\left(\frac{\partial^2 \psi}{\partial x^2} + \frac{\partial^2 \psi}{\partial y^2}\right) = \left(\frac{\varepsilon_1 \omega_0^2}{c^2} - \frac{\partial^2 \varepsilon_2}{\partial z^2} + \frac{1}{\varepsilon_2}\left(\frac{\partial \varepsilon_2}{\partial z}\right)^2 - \frac{3}{4\varepsilon_1^2}\left(\frac{\partial \varepsilon_1}{\partial z}\right)^2 + \frac{1}{2\varepsilon_1}\frac{\partial^2 \varepsilon_1}{\partial z^2}\right)\psi$$

$$(9)$$

For simplicity, let us consider the case of constant $\varepsilon_1$. In such a case z-derivatives of $\varepsilon_1$ disappear, while z-derivatives of $\varepsilon_2$ lead to a small correction to the effective mass. This is indeed the case in the $\varepsilon_2(z) \to 0$ limit if $\varepsilon_2$ approaches zero no slower than $\sim(z-z_{bc})^2$, where the emulated Big Crunch point is located at $z=z_{bc}$. As a result, no light ray can escape the Big Crunch point in such an effective optical spacetime. While the flat space case appears to be the most simple and obvious to analyze, we should point out that other possible geometries of $d\Sigma$, such as the constant negative and constant positive curvature FLRW metrics may be emulated in a similar fashion. In fact, the Big Crunch light concentrators may have any desired shape defined by an arbitrary choice of $d\Sigma^2$. While such designs only work for extraordinary light, making the front surface of the light concentrator rough will make it "darker than black" [17]. Due to broadband divergence of the photonic density of states in hyperbolic metamaterials [14], roughened surface of a hyperbolic metamaterial scatters light preferentially inside the medium into extraordinary photon states, resulting in almost zero (darker than black) reflectance. As demonstrated above, these extraordinary photons will be concentrated into the Big Crunch point. Thus, close to 100% omnidirectional light concentration will be achieved for both polarization states of light.

Detailed analysis of the spherical concentrator design appears to be interesting for another important reason. Such a spherical Big Crunch concentrator is closely related to the hyperlens-based designs of super-resolution optical microscopes [9-13]. Both designs are based on hyperbolic metamaterials having $\varepsilon_1=\varepsilon_\theta=\varepsilon_\phi >0$ and $\varepsilon_2=\varepsilon_r <0$. The latter inequality insures that the radial coordinate behaves as a timelike variable. However, operational requirements of these designs are different. While hyperlens must efficiently transmit super-resolution imaging information from the central area of the lens into the far-field, an efficient Big Crunch concentrator does not need to be an imaging device.

Let us consider the spherical case in more detail, and search for solutions of Maxwell equations (5) in terms of spherical harmonics $Y_l^m(\theta,\phi)$, so that

$$E_r = \psi(r) Y_l^m(\theta,\varphi) e^{-i\omega t} \, , \qquad (10)$$

and $\varepsilon_1=\varepsilon_\theta=\varepsilon_\phi >0$ and $\varepsilon_2=\varepsilon_r <0$ are assumed to depend on $r$. The resulting Maxwell equation for $\psi(r)$ may be written as follows:

$$-\frac{1}{r^2}\frac{\partial}{\partial r}\left(\frac{1}{\varepsilon_1}\frac{\partial}{\partial r}\left(r^2\varepsilon_2\psi\right)\right)+\frac{l(l+1)}{r^2}\psi = \omega^2\varepsilon_2\psi \qquad (11)$$

Solutions of this equation are easier to analyze if an auxiliary radial wave function is introduced as $\chi=\psi\varepsilon_2 r^2$, so that eq.(11) transforms into

$$-\frac{\partial}{\partial r}\left(\frac{1}{\varepsilon_1}\frac{\partial \chi}{\partial r}\right)+\frac{l(l+1)}{\varepsilon_2 r^2}\chi = \omega^2\chi \qquad (12)$$

Assuming for the sake of simplicity that $\varepsilon_1$ and $\varepsilon_2$ are constant, let us analyze solutions of eq.(12) near $r=0$ (near the Big Crunch point). Since $\varepsilon_2<0$, the second term in eq.(12) plays the role of attractive central potential. Analysis of radial Shrodinger equation performed in [18] indicates that depending on the numerical value of $l(l+1)\varepsilon_1/\varepsilon_2$, the photon state may or may not collapse toward the central singularity. Indeed, let us search for solutions of eq.(12) which behave as $\chi\sim r^S$ near r=0. Substitution of $\chi\sim r^S$ into eq.(12), while neglecting slower terms, gives rise to quadratic equation for S:

$$S^2 - S - \frac{\varepsilon_1}{\varepsilon_2}l(l+1) = 0 \quad, \qquad (13)$$

resulting in

$$S = -\frac{1}{2}+\left(\frac{1}{4}+\frac{\varepsilon_1}{\varepsilon_2}l(l+1)\right)^{1/2} \qquad (14)$$

The wave function collapse occurs if

$$l(l+1) > -\frac{\varepsilon_2}{4\varepsilon_1} \qquad (15)$$

Therefore, a radial Big Crunch concentrator is realized for small enough $-\varepsilon_2/\varepsilon_1$ ratios, so that inequality (15) is satisfied for every $l>0$. Obviously, this happens at $-\varepsilon_2/\varepsilon_1<8$. On the other hand, an efficient hyperlens is obtained at large enough $-\varepsilon_2/\varepsilon_1$ ratios. In order to reach high enough $2\pi/l$ angular resolution, the $-\varepsilon_2/\varepsilon_1$ ratio of the hyperlens must exceed *4l(l+1)*. As a result, spatial information with $2\pi/l$ angular resolution will be transmitted into the far field.

Our numerical simulations of light propagation though various Big Crunch concentrators shown in Fig.2 have been performed using COMSOL Multiphysics 4.2a solver. In order to simplify calculations we have considered various two-dimensional distributions of metamaterial parameters $\varepsilon_r$ and $\varepsilon_\phi$, which lead to the same effective attractive potential $l(l+1)\varepsilon_1/\varepsilon_2 r^2$ as in the 3D case. While, this method may not be considered accurate in the $r \rightarrow 0$ limit, representation of electromagnetic properties of the structure itself with effective hyperbolic values of $\varepsilon_1$ and $\varepsilon_2$ also breaks down near the center of concentrator. Despite this difficulty, efficient light concentration near the center of metamaterial patterns has been observed. Moreover, pronounced geometrical shadow behind the structures indicate efficient wide field of view performance. Our simulations, while broadly compatible with the simulation results presented in [1], indicate some essential advantages of the Big Crunch approach compared to the omni-directional concentrators based on the electromagnetic black hole geometries. Unlike the latter, the Big Crunch geometries do not need divergent values of dielectric permittivity near the center of metamaterial structure. Moreover, as demonstrated in Fig.2(c), the Big Crunch geometry may accommodate ellipsoidal shape $d\Sigma^2$ of the concentrator. On the other hand, potential disadvantages of the hyperbolic metamaterial approach include relatively large metamaterial losses.

While experimental demonstration of the Big Crunch concentrator with 3D metamaterials would require sophisticated nanofabrication, experimental demonstration of this concept using PMMA-based plasmonic metamaterials described in detail in refs.[11,15] may be achieved by much simpler means. Our approach is based on similarities between the layered 3D hyperbolic metamaterials and plasmonic hyperbolic metamaterials based on the PMMA stripes formed on the surface of thin gold film [11,15]. Around free space wavelength $\lambda_0 \sim 500$ nm plasmons perceive PMMA stripes on gold as if they are "metallic layers", while gold/vacuum portions of the interface are perceived as "dielectric layers". Thus, at these frequencies plasmons perceive a pattern of PMMA stripes on gold as a layered hyperbolic metamaterial. Rigorous theoretical description of the PMMA-based plasmonic metamaterials developed in ref. [19] confirms this similarity. Fabrication of such 2D plasmonic hyperbolic metamaterial requires only very simple and common lithographic techniques. The required concentric structures were defined using a Raith E-line electron beam lithography system with ~70 nm spatial resolution and developed using a 3:1 IPA/MIBK solution obtained from Microchem. An example of such a structure is shown in the inset in Fig.3(b). The fabricated structures were studied using an optical microscope under illumination with P-polarized 488 nm Argon ion laser, as described in [11,15]. Illumination angle was varied in order to achieve phase-matched excitation of plasmons by the concentric ring grating. Our experimental data are presented in Fig.3. Both circular (a) and elliptic (b,c) concentric metamaterial patterns were fabricated and tested. The 488 nm illumination was from a 45 degree elevation incident from the bottom side of the image going across the metamaterial patterns. The images shown in Fig.3 were formed by plasmon scattering into light and captured by a CCD camera mounted onto the microscope. In agreement with our theoretical modeling described above, enhanced light intensity near the center of plasmonic Big Crunch concentrators is indeed clearly visible in all cases.

In conclusion, we have proposed a gravity-inspired design of a broadband omnidirectional light concentrator based on the cosmological Big Crunch solutions. By mimicking the Big Crunch spacetime via corresponding effective optical metric we make sure that every photon world line terminates in a single point. The proof of principle Big Crunch concentrator designs have been demonstrated in experiments performed with 2D plasmonic metamaterial structures.

**Figure Captions**

**Fig. 1**. Schematic view of the Big Crunch spacetime geometry. All the possible world lines terminate at the Big Crunch point. The goal of our optical concentrator design is to create similar effective "optical spacetime" for photons, so that every photon trajectory terminates at the effective Big Crunch point.

**Fig. 2**. Numerical modeling of light collection at the center of Big Crunch-based omnidirectional light concentrators performed using COMSOL Multiphysics 4.2a solver. Different scale factor $a(r)$ behavior as a function of time-like radial variable was used in frames (a) and (b), while $d\Sigma^2$ had spherical topology in both cases. The topology of $d\Sigma^2$ had been changed to elliptical in frame (c). $\lambda=0.1$ has been chosen in all cases.

**Fig. 3**. Microscopic images of various plasmonic metamaterial structures illuminated with 488 nm laser light show evidence of plasmon concentration. In all cases the illumination direction is from the bottom of each image. The inset in (b) shows AFM image of the example of plasmonic metamaterial structure.

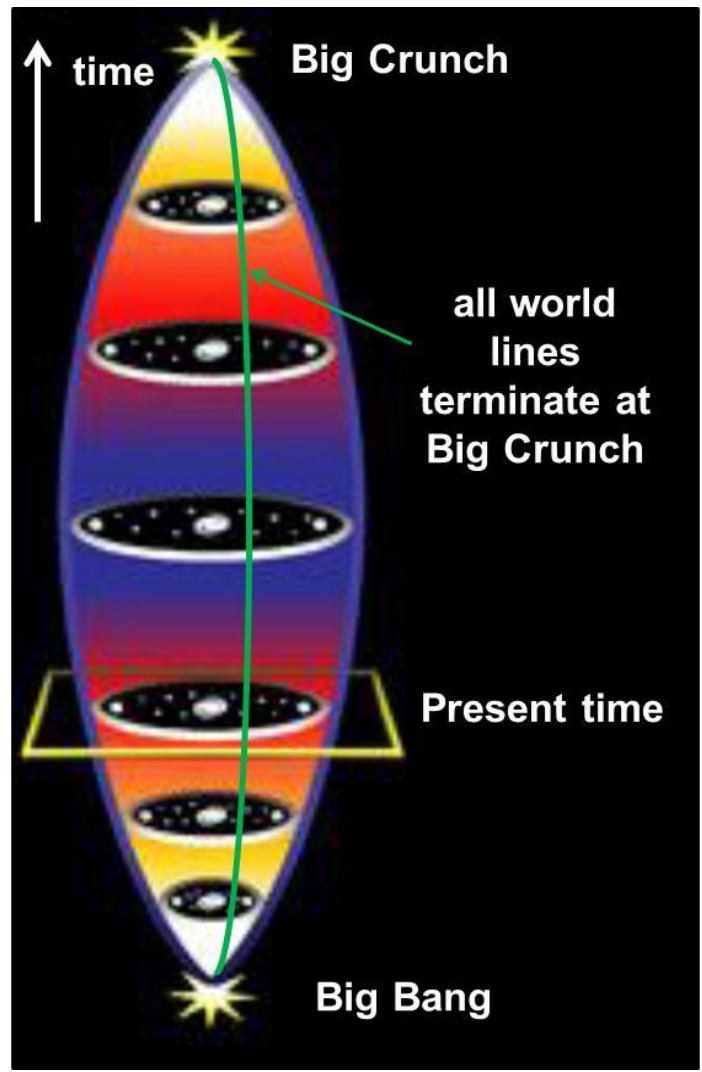

Fig. 1

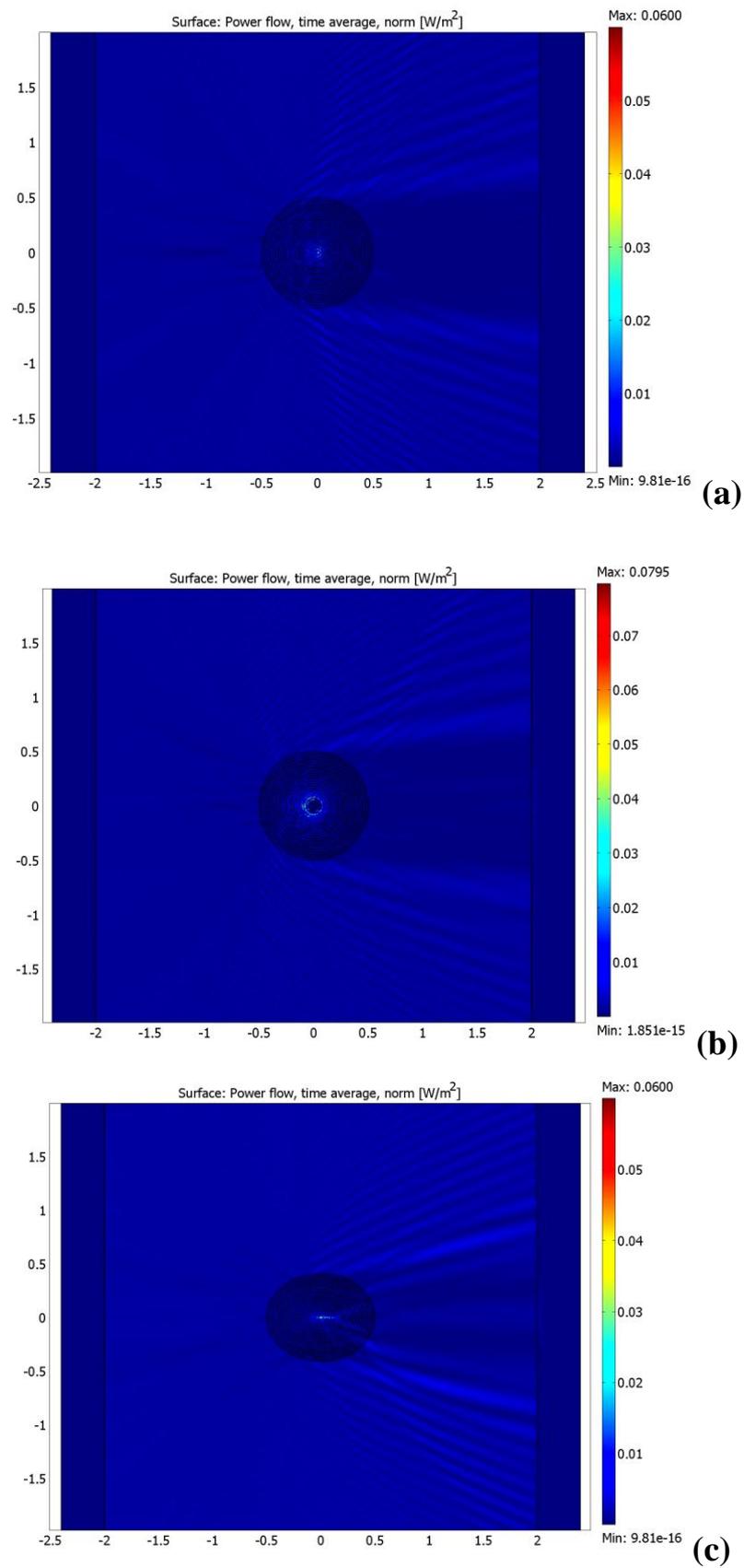

Fig. 2

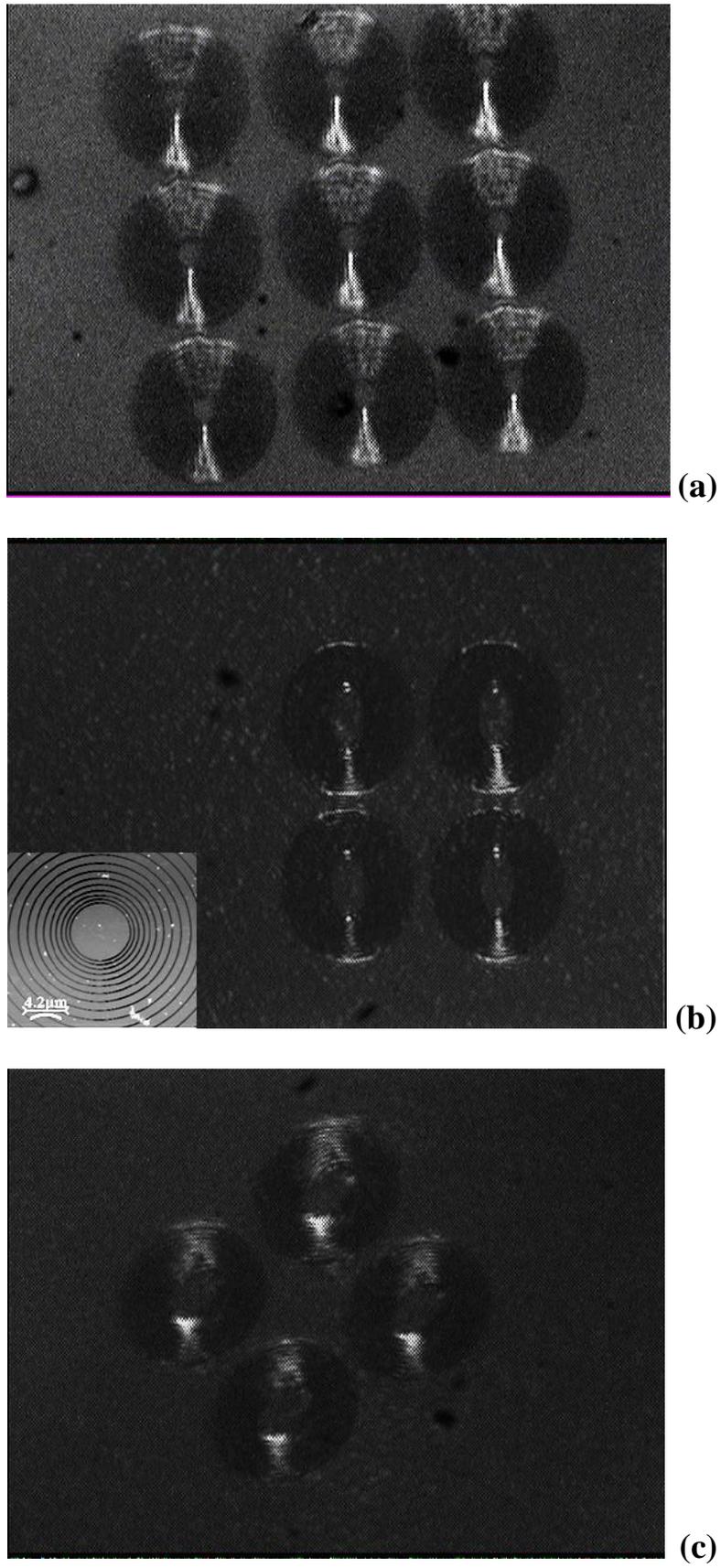

Fig. 3